\documentclass[10pt,conference]{IEEEtran}
\usepackage{amsmath,amsfonts}
\usepackage{algorithmic}
\usepackage{algorithm}
\usepackage{array}
\usepackage[caption=false,font=normalsize,labelfont=sf,textfont=sf]{subfig}
\usepackage{textcomp}
\usepackage{stfloats}
\usepackage{url}
\usepackage{verbatim}
\usepackage{graphicx}
\usepackage{cite}
\usepackage{authblk} 

\begin{document}

\title{On the Trade-Off Between Sum-Rate and Energy Efficiency through the Convergence of HAPS and Active RIS Technologies}

\author[1]{Bilal Karaman}
\author[1]{Ilhan Basturk}
\author[2,3]{Ferdi Kara}
\author[4]{Metin Ozturk}
\author[1]{Sezai Taskin}
\author[5]{Halim Yanikomeroglu}

\affil[1]{\textit{Manisa Celal Bayar University, Manisa, Türkiye}} 

\affil[2]{\textit{KTH Royal Institute of Technology, Sweden}}

\affil[3]{\textit{Zonguldak Bulent Ecevit University, Zonguldak, Türkiye}}

\affil[4]{\textit{Ankara Yıldırım Beyazıt University, Ankara, Türkiye}}

\affil[5]{\textit{Non-Terrestrial Networks Lab., Carleton University, Ottawa, ON, Canada}}




\maketitle

\begin{abstract}

This paper investigates the integration of active reconfigurable intelligent surfaces (RIS) relay with high-altitude platform stations (HAPS) to enhance non-terrestrial network (NTN) performance in next-generation wireless systems. 
While prior studies focused on passive RIS architectures, the severe path loss and double fading in long-distance HAPS links make active RIS a more suitable alternative due to its inherent signal amplification capabilities. 
We formulate a sum-rate maximization problem to jointly optimize power allocation and RIS element assignment for ground user equipments (UEs) supported by a HAPS-based active RIS-assisted communication system. To reduce power consumption and hardware complexity, several sub-connected active RIS architectures are also explored. Simulation results reveal that active RIS configurations significantly outperform passive RIS in terms of quality of service (QoS). Moreover, although fully-connected architectures achieve the highest throughput, sub-connected schemes demonstrate superior energy efficiency under practical power constraints. These findings highlight the potential of active RIS-enabled HAPS systems to meet the growing demands of beyond-cellular coverage and green networking.

\end{abstract}


\section{Introduction}
With the rapid growth in user density and the evolution of Internet of things (IoT) technologies, next-generation wireless networks face increasing demand for high-capacity and always-on connectivity \cite{6Groad}. While densifying terrestrial networks by deploying more base stations (BSs) may seem a viable solution, it is often impractical due to high capital and operational expenditures \cite{kement}. As a result, non-terrestrial networks (NTN) have emerged as a strong candidate to complement terrestrial networks and meet the connectivity needs of future communication systems~\cite{gunessurvey}. 
As a key player of NTNs, high-altitude platform stations (HAPS) can be envisioned as flying solar-powered platforms, such as balloons or aircraft systems, operating around 20 km above the ground. Positioned between terrestrial BSs and low-Earth orbit (LEO) satellites, HAPS offer a balance between coverage, latency, and deployment complexity. 
Additionally, unlike LEO satellites that require frequent handovers due to constant motion, HAPS maintain a quasi-stationary position, enabling stable and persistent connectivity~\cite{gunessurvey}.

HAPS systems offer several key advantages in next-generation communication systems. They can enable high-speed, low-latency connectivity in remote or underserved areas where conventional infrastructure is lacking. Their deployment is significantly faster than that of terrestrial alternatives, such as fiber optic networks or cellular towers. Moreover, HAPS is particularly well-suited for emergency scenarios and provide reliable coverage in geographically challenging or inaccessible regions. While HAPS is often associated with rural deployments, its integration into urban environments as part of multi-tier sixth-generation of cellular network technology~(6G) architectures holds significant promise~\cite{gunessurvey}. 
Rather than functioning as standalone systems, they are expected to complement terrestrial and non-terrestrial networks~\cite{safwanlinkbudget}. 

On the other hand, reconfigurable intelligent surfaces (RIS) are emerging as a key enabler for 6G networks. These energy-efficient, programmable metasurfaces can intelligently reflect and manipulate wireless signals in real-time, enhancing link quality \cite{RIS1}. Their extremely low power consumption, ability to boost signal strength, and ease of deployment—even in dense or complex environments—make them ideal for next-generation wireless systems \cite{huang2019reconfigurable}. When integrated with HAPS (i.e., mounting RIS on HAPS), RIS can significantly extend network coverage across both urban and rural areas. This synergy can also reduce infrastructure costs for operators while enabling advanced applications, such as precision agriculture, emergency communications, and aerial delivery systems. Furthermore, the combined use of HAPS and RIS supports sustainable, green networking objectives by lowering energy consumption and improving spectrum efficiency.

Mounting RIS on aerial platforms offers superior system performance and improved link budget efficiency. This approach benefits from essential advantages, including strong line-of-sight~(LoS) connectivity, enhanced channel conditions, broad coverage areas, and flexible positioning capabilities \cite{safwanlinkbudget,safwanmagazine}. Additionally, HAPS can meet its energy needs through solar power and, thanks to their large surface area, can accommodate extensive RIS deployments to achieve significant reflection gains \cite{gunessurvey}. Studies in \cite{safwanGlobecom} and \cite{safwanLetter} showed that HAPS-RIS-enabled beyond-cell communication could effectively complement terrestrial networks, manage the growing user density, and provide service to unconnected users through a dedicated control station (CS). In \cite{safwanGlobecom}, sum-rate maximization, reflecting unit minimization, and minimum rate maximization problems were investigated for the HAPS-RIS-assisted communication system. In \cite{safwanLetter}, a resource efficiency optimization problem was studied to increase the number of connected users in the terrestrial network supported by HAPS-RIS. In \cite{alfattani2022multi}, a backhaul link was established between a ground station and an isolated BS in a remote area using the HAPS-RIS architecture. Moreover, the performance of the HAPS-RIS system was compared with that of a traditional HAPS-relay system. The results demonstrated that HAPS-RIS outperforms the conventional relay system in terms of both sum-rate and energy efficiency. In \cite{azizi2024exploring}, a novel architecture was proposed in which HAPS-RIS and UAV-BS systems operate in tandem. In their proposed system, users are served by defining a HAPS-RIS zone and a UAV zone on the ground. The study in \cite{deka2024performance} explored free-space optics (FSO)-based relaying system between ships using a HAPS-integrated RIS.

However, all these studies in the literature focused on passive RIS systems. Considering the distances between the CS and the HAPS, as well as between the HAPS and the user equipments (UEs), passive RIS systems become significantly disadvantaged due to the multiplicative fading effects of the two-way communication channel. By integrating the active RIS concept into the HAPS, the incoming signal can be amplified before reflection, thereby enhancing both the spectral and energy efficiency of the system. Therefore, in this study, we propose a HAPS-assisted communication system empowered by active RIS technology. To the best of our knowledge, this is the first study that incorporates active RIS into HAPS-assisted communication systems. The contributions of this paper are as follows:

\begin{itemize}

    \item We formulate a sum-rate maximization problem that aims to determine the optimal power and RIS element allocation strategies for UEs supported by the HAPS-active RIS-assisted communication system.
    
    \item Considering the large number of RIS elements on the HAPS, we evaluate three different sub-connected active RIS architectures in addition to the fully-connected one. We also analyze their energy efficiency performances.
    

    \item Through numerical analysis, the trade-off between throughput and energy efficiency is examined by comparing the sub-connected active RIS deployment with its fully-connected counterpart and the passive RIS scheme.
\end{itemize}

The remainder of this study is organized as follows. Section II outlines the system model, while Section III details the problem formulations. In Section IV, we discuss the proposed solution for solving the optimization problem. In Section V, we present the energy efficiency analysis, followed by presenting the numerical results and discussions in Section VI. Finally, Section VII concludes the study.

\section{System Model}

We consider an urban communication scenario comprising \( K \) UEs, \( D \) terrestrial BSs, a single HAPS-RIS system, and one CS, as shown in Fig.~1(a). Due to typical urban impairments, such as shadowing and blockages, most UEs experience non-line-of-sight (NLoS) conditions. Based on the link quality and serving capability of the BSs, a subset \( \mathcal{K}_1 = \{1, \dots, K_1\} \) is served directly by the terrestrial BSs (referred to as \textit{within-cell} users~\cite{safwanGlobecom}).

The remaining UEs, denoted by \( \mathcal{K}_2 = \{1, \dots, K_2\} \), which lack direct connectivity with the terrestrial BSs, or can not exceed the minimum data rate requirements ($R_\text{min}$) are served by the CS via the HAPS-RIS (we refer to them as \textit{HAPS-RIS} users). The CS is assumed to be located within the coverage area of the HAPS. The overall set of UEs is expressed as \( \mathcal{K} = \mathcal{K}_1 \cup \mathcal{K}_2 \). Building upon the system model presented in \cite{safwanGlobecom}, we further extend the framework by considering a larger geographical area with an increased number of UEs. Accordingly, \textit{within-cell} and \textit{HAPS-RIS} users, as well as terrestrial BSs and CS, are illustrated in Fig.~1(b).

We assume that the CS serves the UEs in \( \mathcal{K}_2 \) using orthogonal subcarriers to avoid interference in the NTN group. Additionally, \textit{within-cell} and \textit{HAPS-RIS}-assisted communications are allocated to distinct frequency bands, while maintaining equal subcarrier bandwidth \( B_{\mathrm{UE}} \). This design ensures no interference between the \textit{within-cell} and \textit{HAPS-RIS}-assisted communications groups. In this study, the focus is placed on the UEs in set \( \mathcal{K}_2 \) to reveal the potential of the HAPS-based active RIS communication system.


\begin{figure*}
    \centering
    \subfloat[]{\includegraphics[width=0.35\textwidth]{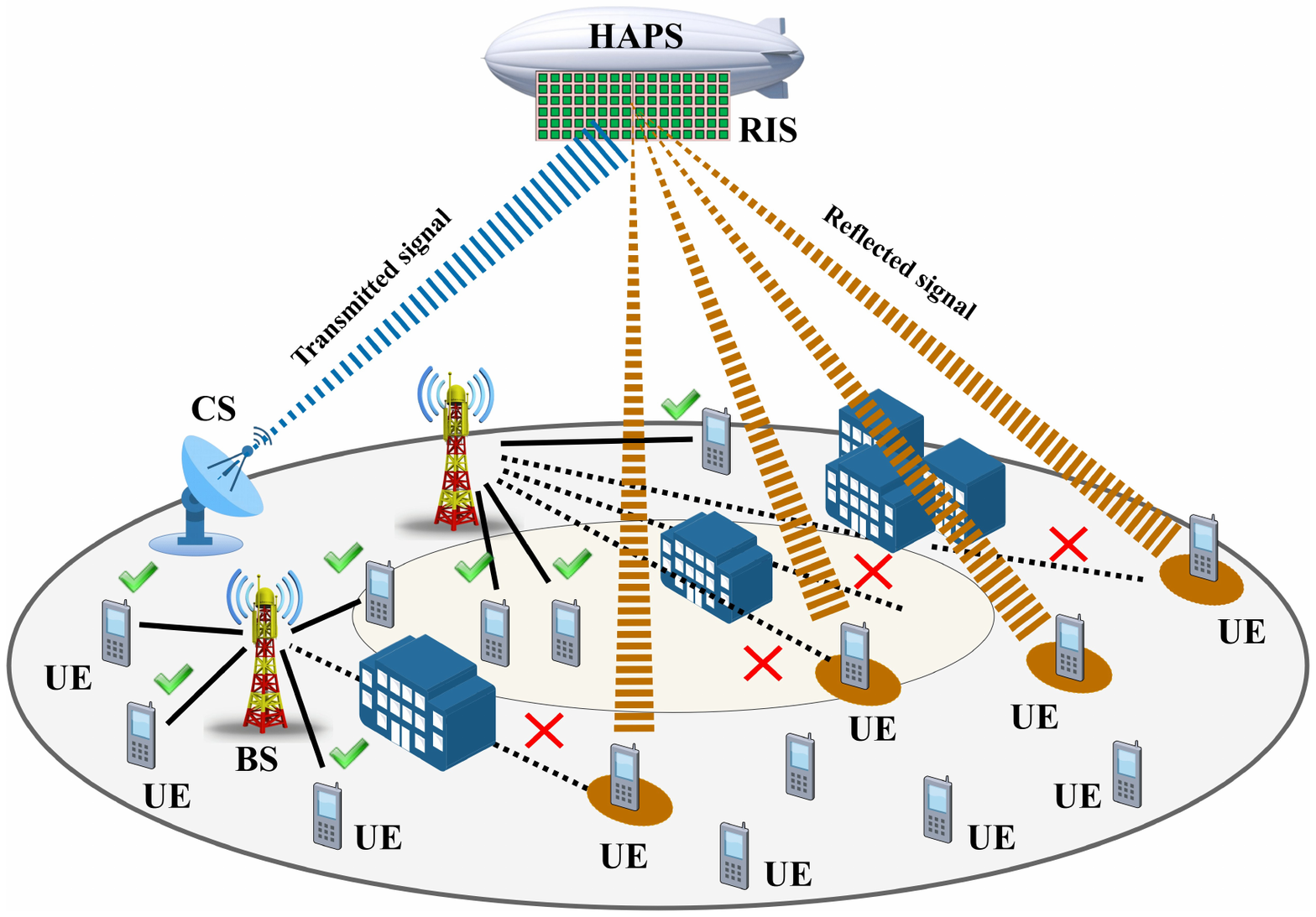}} 
    \subfloat[]{\includegraphics[width=0.35\textwidth]{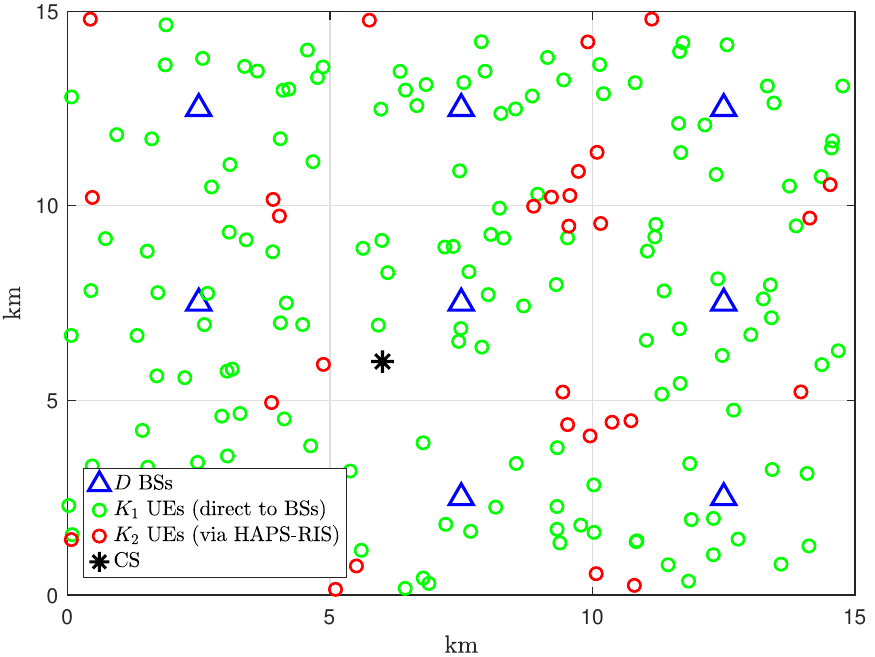}} 
    \subfloat[]{\includegraphics[width=0.30\textwidth]{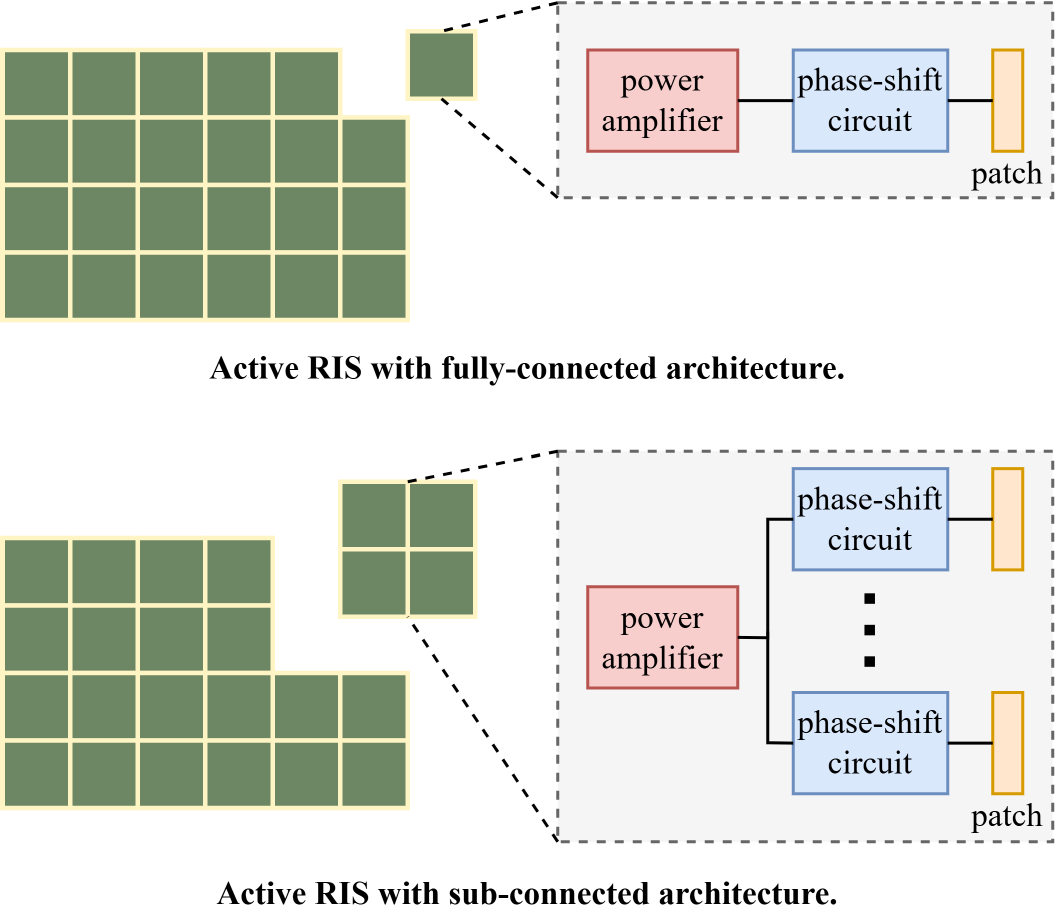}}
   
    \caption{(a)  System model for HAPS-RIS assisting communications.  (b)  The considered system model in a 15 km x 15 km area with 1 HAPS-RIS, 1 CS, 9 terrestrial networks BSs and 180 UEs. Green nodes are served by terrestrial networks BSs, red nodes are served by HAPS-RIS.  (c) Comparison between fully-connected and sub-connected architectures for active RIS. }
\end{figure*}

\subsection{Active RIS Architecture}

The active RIS has a total of $N$ reflecting elements, introducing a controllable phase shift after amplification. To reduce the power consumption and complexity of the RIS, a sub-connected architecture is employed~\cite{liu2021active}. Unlike conventional RIS, the active RIS enables a group of $T$ elements to share a common amplifier with a unified amplification factor. However, each element still maintains its own phase shift control. The comparison between
fully-connected and sub-connected architectures for active RIS is given in Fig.1(c).

Also, $Q = N / T$ defines the required number of amplifiers for this system.
The precoding matrix $\boldsymbol{\Phi} \in \mathbb{C}^{N \times N}$ at the active RIS can be expressed as $\mathrm{diag}(\boldsymbol{\Phi}) = \mathrm{diag}(\boldsymbol{\Theta} \boldsymbol{\Gamma} \mathbf{\beta})$, where $\boldsymbol{\Theta} \in \mathbb{C}^{N \times N}$ denotes the diagonal matrix of phase shifts, similar to that used in passive RIS architectures. The vector $\mathbf{\beta} \in \mathbb{R}_+^{Q \times 1}$ represents the amplification gains applied at the RIS elements. Furthermore, $\boldsymbol{\Gamma} \in \mathbb{C}^{N \times Q}$ is the indicator matrix that indicates the connection between power amplifiers~(PAs) and phase shift circuits. Without loss of generality, we assume $\boldsymbol{\Gamma} = \mathbf{I}_Q \otimes \mathbf{1}_T$.

The gain of the active RIS can be defined as follows ~\cite{yigit2022hybrid}:

\begin{equation}
\beta\le \sqrt{\frac{P_{\text{A}}}{P_{\text{t}}\left\| h \right\|^{2}+\sigma_z^{2}}},
\end{equation}
where $P_\text{t}$ is the transmission power of CS, $h \in \mathbb{C}^{N \times 1}$ is the channel vector between CS and RIS elements, $P_\text{A}$ is the power of active RIS, and $\sigma_z^2$ is the dynamic noise power.

\subsection{Active RIS-assisted Communication}

In this study, UEs that are either unable to establish a direct link with the terrestrial BS or cannot achieve the minimum required data rate (i.e., $R_\text{min}$) are supported by the CS via the HAPS-RIS architecture. It is assumed that the CS serves the UEs in set \( \mathcal{K}_2 \) using the orthogonal frequency-division multiplexing (OFDM) protocol, thereby eliminating inter-UE interference. Additionally, both \textit{within-cell} and \textit{HAPS-RIS}-assisted transmissions are allocated to distinct, orthogonal frequency bands, with an identical bandwidth \( B_{\text{UE}} \) assigned to each. As a result, interference between \textit{within-cell} UEs in \( \mathcal{K}_1 \) and \textit{HAPS-RIS} UEs in \( \mathcal{K}_2 \) is completely avoided. Based on this, the received signal at UE \( k \in \mathcal{K}_2 \) can be given by
\begin{equation}
\label{eq_received_signal}
y_{k}=\sqrt{P_{k}}\left(\boldsymbol{g}_{k}^H\boldsymbol{\Phi_{k}}\boldsymbol{h}_{k}\right)
x_{k} + \boldsymbol{g}_{k}^H\boldsymbol{\Phi_{k}}z_k + n_k,
\end{equation}
where $P_{k}$ and $x_{k}$ represent the transmit power and signal of UE $k$, $z_{k}$ denotes the dynamic noise with zero mean and variance $\sigma_{z}^{2}$, caused by the active RIS amplifiers, and $n_{k}$ denotes the additive white Gaussian noise (AWGN) with zero mean and variance $\sigma^{2}$. $\boldsymbol{h}_{k} \in\mathbb{C}^{
N_k\times1}$  is the vector containing the effective channel coefficients between CS and each reflecting element at the HAPS-RIS in which ${N_{k}}$ indicates the total number of RIS units allocated to the UE $k$. Likewise, and $\boldsymbol{g}_{k} \in\mathbb{C}^{
N_k\times1}$ is the vector containing the effective channel coefficients between each element at the HAPS-RIS and UE $k$\footnote{We assume negligible channel variations between the UE and RIS elements due to dominant LoS links.}. In both, each entry is defined as $h_{k}=\sqrt{\frac{G_\text{T}}{{L}_{\text{CH}}}}e^{-j\theta_{i}}$ and $g_{k}=\sqrt{\frac{G_\text{R}}{{L}_{\text{HU}_{k}}}}e^{-j\psi_{i}}$, $i=1,2,\dots, N_k$ respectively, where $G_{\text{T}}$ is the antenna gain of the control station, $G_{\text{R}}$ is the receiver antenna gain, ${L}_{\text{CH}}$ denotes the path loss from the CS to the HAPS-RIS, while ${L}_{\text{HU}_{k}}$ is path loss from the HAPS-RIS to UE $k$. $\theta_i \sim [0, 2\pi]$ and $\psi_i \sim [0, 2\pi]$ denote the phase of the channels between CS-HAPS and HAPS-RIS, respectively. 
$\boldsymbol{\Phi_{k}}\in\mathbb{C}^{N_k \times N_k}$ is a diagonal phase shift matrix at the RIS where $\textbf{diag}\{\boldsymbol{\Phi_{k}}\}=[\beta_1 e^{j\phi_1},\ \beta_2 e^{j\phi_2}, \dots,\beta_i  e^{j\phi_i}, \dots, \beta_{N_k} e^{j\phi_{N_k}}]$ and optimally $\phi_i=\theta_i+\psi_i$. For the sake of simplicity, we consider that all active reflecting elements exhibit an identical reflection gain, i.e., $\beta_i = \beta \  \text{for } \forall i \in \{1, 2, \ldots, N_k\}$.

Using (2), the signal-to-noise ratio (SNR) at UE $k$ can be expressed as

\begin{equation}
\label{eq_SNR}
\gamma_{k}=\frac{P_{k} \left| \boldsymbol{g}_{k}^T\boldsymbol{\Phi_{k}}\boldsymbol{h}_{k} \right|^{2}}{\left\| \boldsymbol{g}_{k}^{T}\boldsymbol{\Phi}_{k} \right\|^2 \sigma_{z}^{2} + N_0B_{\text{UE}}},
\end{equation}
and the corresponding achievable rate can be expressed as

\begin{equation}
\label{eq_sum_rate}
R_{k} = B_{\text{UE}}\log_2(1+\gamma_{k}).
\end{equation}

\subsection{3GPP Path-Loss Model}

The path loss for LoS and NLoS conditions, for the link CS $\to$ HAPS-RIS, denoted by ${L}_{\text{CH}}$, can be written as \cite{3GPP_UE}

\begin{equation}
\label{eq_CStoHAPS_PL}
L_{\text{CH}} = \mathcal{P}_{\text{CH}}^{\text{LoS}}L_{\text{CH}}^{\text{LoS}}+\mathcal{P}_{\text{CH}}^{\text{NLoS}}L_{\text{CH}}^{\text{NLoS}},
\end{equation}
and for HAPS-RIS $\to$ UE $k$ link is given by
\begin{equation}
\label{eq_CStoHAPS_PL}
L_{\text{HU}_{k}} = \mathcal{P}_{\text{HU}_{k}}^{\text{LoS}}L_{\text{HU}_{k}}^{\text{LoS}}+\mathcal{P}_{\text{HU}_{k}}^{\text{NLoS}}L_{\text{HU}_{k}}^{\text{NLoS}},
\end{equation}
where $\mathcal{P}_{\text{CH}}^{\iota}=c_{1}\vartheta_{1}^{c_{2}}+c_{3}$, $\mathcal{P}_{\text{HU}_{k}}^{\iota}=c_{1}\vartheta_{2}^{c_{2}}+c_{3}$, $L_{\text{CH}}^{\iota}=L_{\text{b}}^{\iota}+L_{\text{g}}+L_{\text{s}}+L_{\text{e}}$, $L_{\text{HU}_{k}}^{\iota}=L_{\text{b}}^{\iota}+L_{\text{g}}+L_{\text{s}}+L_{\text{e}}$, $L_{\text{b}}^{\iota}=FSPL+CL^{\iota}+X^{\iota}$, $FSPL=32.45+20\log_{10}(f)+20\log_{10}(d_{\text{3D}})$, $d_{\text{3D}}=\sqrt{R_{\text{E}}^{2}\sin^{2}(\vartheta)+H_{\text{z}}^{2}+2H_{\text{z}}R_{\text{E}}}-R_{\text{E}}\sin(\vartheta)$, $\iota=\{ \text{LoS, NLoS} \}$, $\mathcal{P}_{\text{CH}}^{\iota}$ and $\mathcal{P}_{\text{HU}_{k}}$ are the LoS probability functions, $L_{\text{CH}}^{\iota}$ and $L_{\text{HU}_{k}}^{\iota}$ are the path losses, $c_{1}$, $c_{2}$, and $c_{3}$ are the parameters that depend on the urban environment, which are articulated in \cite{safwanlinkbudget}, $\vartheta_{1}$
 and $\vartheta_{2}$ are the elevation angles,  $L_{\text{b}}^{\iota}$ is the basic path loss, $L_{\text{g}}$ is the attenuation
 due to atmospheric gasses, $L_{\text{s}}$ is the attenuation due to either ionospheric or tropospheric scintillation, and $L_{\text{e}}$ is
 the building entry loss, expressed in dB. Also, $L_{\text{b}}^{\iota}$ represents the signal's free-space propagation loss (FSPL), clutter loss ($CL^{\iota}$), and shadow fading ($X^{\iota}$). In FSPL,  $f_{\text{c}}$ is the operation frequency, $d_{\text{3D}}$ is the 3D distances between the HAPS and the terrestrial UEs or HAPS and the CS, expressed as a function of the HAPS altitude $H_{\text{z}}$, and $R_{\text{E}}$ denotes the Earth’s radius. 

\section{Sum Rate Maximization Problem Formulation}

In this section, we explore a sum-rate maximization problem for UEs supported by HAPS-RIS system, focusing on the total power at the CS and the total RIS elements at the HAPS. Therefore, the proposed HAPS-active RIS system can enhance the QoS requirements of unserved UEs in the terrestrial network. 
The objective of this problem is to maximize the sum rates of the UEs in the $\mathcal{K}_{2}$ set by optimally allocating system resources such as transmit power and reflecting units. Therefore, the optimization problem formulation is given by
\begin{subequations}
\begin{align}
\label{eq_optimization}
& \max_{\Phi_{k}, N_{k}, P_{k}}  \sum_{k=1}^{K_{2}}R_{k}  ,
\\& \quad
\text{s.t:}
\ \ R_{k}\ge R_{\text{th}}, \ \forall k=1,...,K_{2},
\\&  \ \ \ \ \ \ \ \  \sum_{k=1}^{K_{2}}N_{k} \le N_{\text{max}}, 
\\& \ \ \ \ \ \ \ \ \ \phi_{i} \in \left\{ 0,\Delta\phi,2\Delta\phi, ...,(2^{b}-1)\Delta\phi  \right\}, \notag \\& \ \ \ \ \ \ \ \ \ \ \ \ \ \ \ \ \ \ \ \ \ \ \ \ \ \ \ \ \ \ \ \ \ \ \ \ \ \ \ \ \ \ \  \forall i= 1,...,N_{k},
\\& \ \ \ \ \ \  \ \ \sum_{k=1}^{K}P_{k}\le P_{\text{max}}, \forall k=1,...,K_{2},
\\& \ \ \ \ \ \ \ \ N_{k} \in \left\{ N_{k,\text{min}}, N_{k,\text{min}}+1, ..., N_{k,\text{max}} \right\}, \notag
\\& \ \ \ \ \ \ \ \ \ \ \ \ \ \ \ \ \ \ \ \ \ \ \ \ \ \ \ \ \ \ \ \ \ \ \ \ \ \ \ \ \ \ \ \forall k=1,...,K_{2},
\\& \ \ \ \ \  \ \ \ P_{k,\text{min}}\le P_{k}\le P_{k,\text{max}}, \forall k=1,...,K_{2},  
\\& \ \ \ \ \ \ \ \ T \in \left\{ 1, 512, 1024, 2048 \right\},
\end{align}
\end{subequations}
where (7b) guarantees that each UE meets the minimum required rate. Constraint (7c) ensures that the total reflecting units allocated to all UEs do not exceed the maximum number available at the HAPS. $N_{\text{max}}$ is limited depending on the size of the HAPS in practice. Constraint (7d) defines the discrete range of phase shifts for the reflectors, $\Delta\phi_{i,m} = 2\pi / 2^{b}$ with representing the uniform quantization of phase shifts using $b$ bits per RIS element. Constraint (7e) guarantees that the total allocated power for each UE does not surpass the maximum power, $P_{\text{max}}$, available from the CS. Constraints (7f)-(7g) ensure a practical and equitable allocation of both reflecting units and CS power, respectively, where $N_{k,\text{min}}, N_{k,\text{max}}, P_{k,\text{min}}, P_{k,\text{max}}$  denote the minimum and maximum number of allocated RIS elements and power per UE, respectively. Constraint (7h) defines the sub-connected active RIS architecture. For benchmarking, the proposed active RIS system is compared with four schemes:
\begin{itemize}
    \item \textbf{Scheme I:} assigns one PA per RIS element.
    \item \textbf{Scheme II:} assigns one PA per 512 RIS elements.
    \item \textbf{Scheme III:} assigns one PA per 1024 RIS elements.
    \item \textbf{Scheme IV:} assigns one PA per 2048 RIS elements.
\end{itemize}

\section{Proposed Solution}

In this section, we present the solution methodology for the previously defined sum-rate maximization problem. Since $N_{k}$ and $\Phi_{k}$ are discrete variables, solving problem (7) optimally in polynomial time is quite challenging. Furthermore, $\Phi_{k}$ in (7d) typically has a range of discrete phase shifts. Nonetheless, numerous studies have shown that the performance gap between continuous phase shifts and discrete phase shifts is approximately 30\% \cite{discreteLoss}. Considering the large number of RIS elements that can be equipped on the HAPS, the variables $N_{k}$ and $\Phi_{k}$ are relaxed to continuous variables to solve the problem optimally. With this relaxation, the objective function and constraints of problem in (7) transform into posynomials\footnote{`Posynomial' is a function formed by summing positive polynomials [16].}~\cite{boyd2007tutorial}. The final solution to problem in (7) is approximated as $N_{k} \approx \left\lceil N_{k}^{*} \right\rceil$ from the solution of the relaxed problem. Furthermore, for high SNR values, the rate of UE $k$ given in (4) can be approximated as $R_{k}\approx B_{\text{UE}}\log_{2}\left( \gamma_{k} \right)$. Consequently, the sum rate for the set of $\mathcal{K}_{2}$  UEs is given by
\begin{equation}
\label{eq_SNR_sum}
\sum_{k=1}^{K_{2}}R_{k} \approx B_{\text{UE}}\log_{2}\left( \prod_{k=1}^{K_{2}}\gamma_{k} \right).
\end{equation}

The rate constraint (7b) for each UE can be expressed in terms of its SNR. Thus, problem (7) can be reformulated as
\begin{subequations}
\begin{align}
\label{eq_sub_optimization}
& \min_{\Phi_{k}, N_{k}, P_{k}}  \frac{1}{\prod_{k=1}^{K_{2}}\gamma_{k}}  ,
\\& \quad
\text{s.t:}
 \   \frac{1}{\gamma_{k}} \le \frac{1}{\gamma_{\text{min}}}, \ \forall k=1,2,...,K_{2},
\\& \ \ \ \ \ \ \ (7\text{c}) - (7\text{h}).
\end{align}
\end{subequations}

Since both objective and constraints of problem (9) are posynomials, optimal solutions can be efficiently found using a geometric programming approach in polynomial time~\cite{boyd2007tutorial}.

\section{Energy Efficiency Analysis}

The energy efficiency of a HAPS-based active RIS-assisted communication system is defined as the ratio of the total achievable sum rate to the overall power consumption. As demonstrated in \cite{liu2021active}, \cite{khennoufa2024multi}, it can be expressed as
\begin{equation}
E_{m} = \frac{R_{m}}{P_{m}},
\end{equation}
where \( m = \{\text{Scheme I, Scheme II, Scheme III, Scheme IV}\} \), and the analysis focuses on the power consumption associated with the communication payload. The total power consumption is expressed as $P = P_\text{t} + P_\text{s} + N P_{\text{sw}} + N P_{\text{dc}} + QP_\text{A} + K_2 P_k$, where \( P_{\text{t}} \) is the total transmission power, \( P_\text{s} \) denotes the static circuit power at the CS, \( P_{\text{sw}} \) is the power required for phase shifting of the RIS elements, and \( P_{\text{dc}} \) is the biasing power for RIS elements. $N$ denotes the total number of RIS elements, and $Q$ represents the total number of PAs. \( P_\text{A} \) and \( P_k \) refer to the power consumed by each PA and the UE $k$, respectively.

\section{Numerical Results and Discussion}

In this section, we evaluate the performance of the proposed HAPS-RIS-assisted communication approach by comparing the fully-connected and various sub-connected active RIS schemes. The channel parameters used in all simulations for the HAPS-RIS system are given by $c_{1}=9.668$, $c_{2}=0.547$, $c_{3}=-10.58$, $L_{\text{g}}=L_{\text{e}}=10$ dB, $L_{\text{s}}=14.7\vartheta^{(-1.136)}$, $CL^{\text{LoS}}=0$ dB, $CL^{\text{NLoS}}$, $X^{\text{LoS}}$ and $X^{\text{NLoS}}$ are based on [15, Table 6.6.2-2], $H_{\text{z}}=20$ km and $R_{\text{E}}=6378$ km. The HAPS is positioned at \((7500\ \mathrm{m},\ 7500\ \mathrm{m},\ 20000\ \mathrm{m})\), the CS is located at \((6000\ \mathrm{m},\ 6000\ \mathrm{m},\ 0\ \mathrm{m})\), and the ground UEs are randomly distributed within the total area, as shown in Fig.~1(b). The rest of the simulation parameters are listed in Table I.

\begin{table}
\fontsize{8.0pt}{8.0pt}\selectfont
\begin{center}
\caption{Simulation Parameters.}
\renewcommand{\arraystretch}{1.3} 
\begin{tabular}[!h]{|c| c||c| c|}
    \hline
    \textbf{Parameter} & \textbf{Value} & \textbf{Parameter} & \textbf{Value}\\
    \hline
    $K_2$ & 30 & $N_{k,\text{min}}$ & $1000$  \\
    \hline
    $f_\text{c}$ & $2$ GHz & $R_\text{min}$ & $2$ Mbps  \\
    \hline
    $B_{\text{UE}}$ & $2$ MHz & $P_{\text{A}}$ &  $33$ dBm \cite{khennoufa2024multi} \\
    \hline
    $G_{\text{T}}$ & $43.2$ dB  & $P_{\text{sw}}$ & $7.8$ mW \cite{huang2019reconfigurable}  \\
    \hline
    $G_{\text{R}}$ & $0$ dB  & $P_{\text{dc}}$ & $10$ dBm \\
    \hline
    $P_{\text{max}}$ & $33$ dBm  & $P_{\text{s}}$ & $10$ dBm \\
    \hline
    $P_{k,\text{max}}$ & $30$ dBm   & $P_{k}$ & $10$ dBm \\
    \hline
    $P_{k,\text{min}}$  & $5$ dBm & $\sigma_z^2$ & $-80$ dBm  \\
    \hline
    $N_{k,\text{max}}$ &  $50000$ & $N_{0}$ & $-174$ dBm/Hz  \\
    
    \hline
\end{tabular}
\end{center}
\end{table}

In Fig.~2, the relationship between the total number of RIS elements and the overall sum-rate for UEs served by the HAPS-RIS system is illustrated. Passive RIS-assisted communication \cite{safwanGlobecom} is used as a benchmark, as commonly adopted in the literature. In the considered scenario, to ensure that the passive RIS system satisfies the \( R_{\min} \) threshold, the analysis begins with \( N_{\max} = 389{,}120 \) elements and extends up to $716{,}800$ elements. It is worth noting that these element counts correspond to an approximate surface area of \( 350\ \mathrm{m}^2 \) to \( 645\ \mathrm{m}^2 \), assuming each reflector unit has a size of \( (0.2\lambda)^2 \). When the maximum transmission power is set to \( P_{\max} = 33\,\mathrm{dBm} \) and each PA operates at \( \rho = 33\,\mathrm{dBm} \), significant improvements in data rate are observed for both fully- and sub-connected active RIS architectures.

\begin{figure}[!t]
\centering
\includegraphics[width=3.1in]{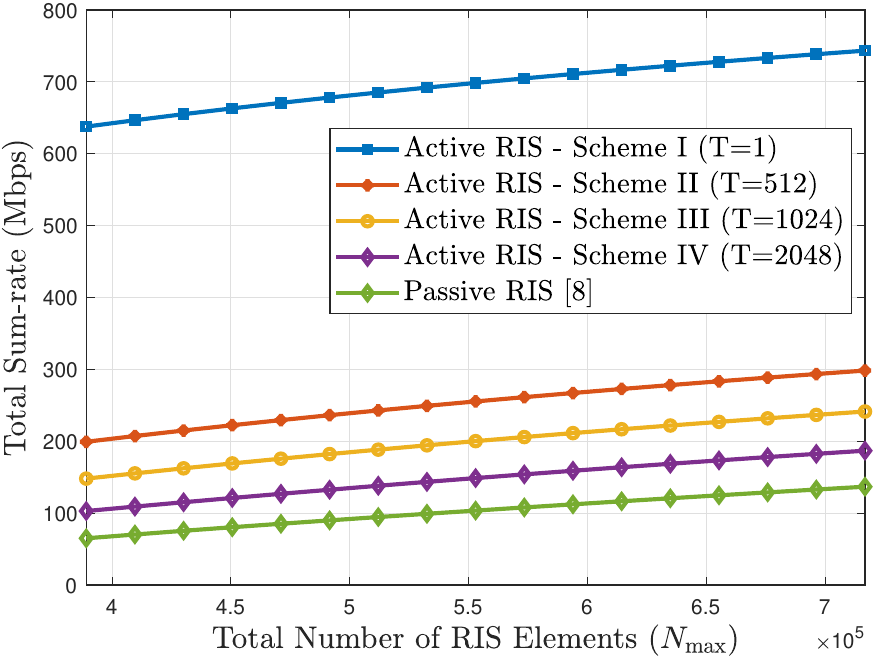}
\caption{Relation between total number of RIS elements and total sum-rate for HAPS-RIS served UEs.}
\label{fig_1_conc}
\end{figure}

In active RIS architectures, the total data rate increases with the number of PAs. However, due to the trade-off between sum-rate and energy efficiency, a large increase in the number of PAs negatively impacts energy efficiency. As shown in Fig.~3, the scenario where each RIS element is connected to a dedicated PA yields the lowest energy efficiency due to high power consumption. Moreover, in sub-connected architectures, as the number of RIS elements—and consequently, the number of PAs—increases, the system exhibits a downward trend in energy efficiency.
\begin{figure}[!t]
\centering
\includegraphics[width=3.1in]{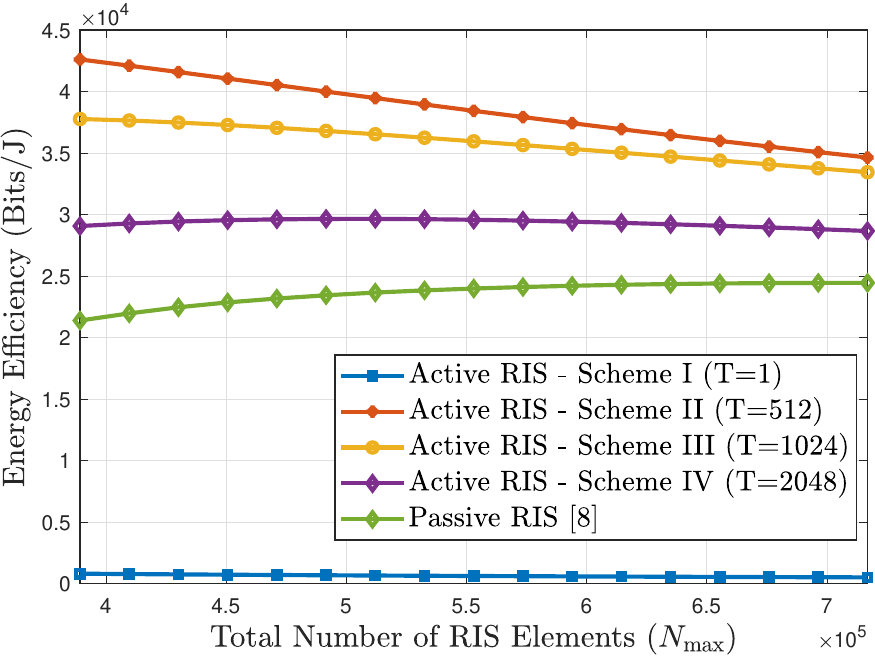}
\caption{Relation between total number of RIS elements and energy efficiency for HAPS-RIS served UEs.}
\label{fig_2_conc}
\end{figure}
Fig.~4 illustrates the impact of active RIS amplification power on the total sum-rate. In this case, the transmission power is fixed at \( P_{\max} = 33\,\mathrm{dBm} \) and the number of RIS elements is set to \( N_{\max} = 389{,}120 \). This allows for a clearer evaluation of the impact of PA output power.
 As the PA output power increases, the total data rate of the system also improves. On the other hand, in sub-connected architectures where each PA is connected to a smaller group of RIS elements, the total number of PAs increases. Consequently, higher energy efficiency is achieved at lower PA power levels in such configurations. As shown in Fig.~5, Scheme~II achieves the highest energy efficiency when \( \rho = 33\,\mathrm{dBm} \). In contrast, since Scheme~IV contains fewer PAs overall, it provides the highest energy efficiency when \( \rho = 40\,\mathrm{dBm} \).

 \begin{figure}[h!]
\centering
\includegraphics[width=3.1in]{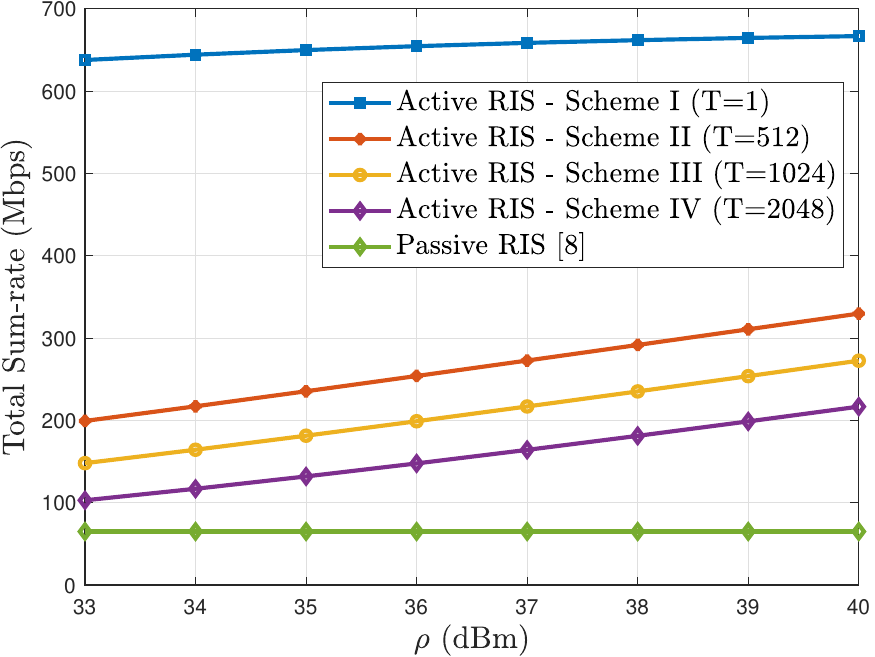}
\caption{Impact of active RIS amplification power on total sum-rate for HAPS-RIS users.}
\label{fig_3_conc}
\end{figure}

In summary, as the main objective of this study, the results obtained through the simulation campaigns highlight the trade-off between performance and energy efficiency (and hardware complexity) in active RIS-aided HAPS systems. While fully-connected architectures maximize the system throughput, they incur higher power consumption due to the large number of PAs, making them less favorable in energy-constrained deployments. On the other hand, sub-connected schemes, particularly those with moderate group sizes (e.g., Scheme~II), offer a more balanced solution by achieving near-optimal sum-rate with significantly improved energy efficiency. This demonstrates the potential of sub-connected active RIS architectures for scalable and sustainable non-terrestrial network designs. Additionally, it is observed that increasing the amplification power beyond a certain point yields diminishing returns in energy efficiency, emphasizing the importance of optimal PA power tuning.

\begin{figure}[!t]
\centering
\includegraphics[width=3.1in]{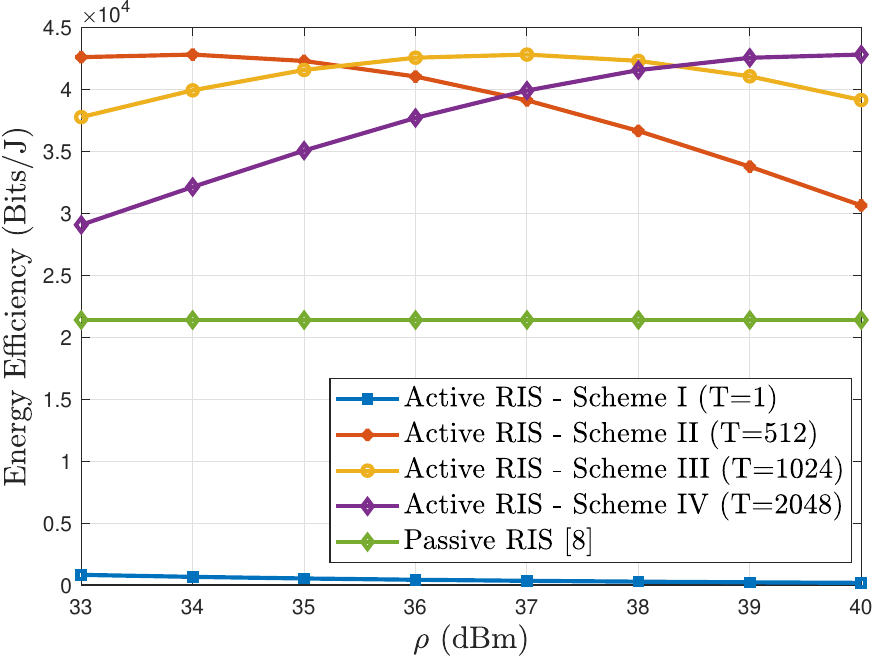}
\caption{Impact of active RIS amplification power on energy efficiency for HAPS-RIS users.}
\label{fig_4_conc}
\end{figure}

\section{Conclusion}

This paper presents a novel HAPS-assisted communication framework empowered by active RIS technology. By addressing the limitations of passive RIS in high-altitude settings, we demonstrate that active RIS integration enables significant improvements in both spectral and energy efficiency. A sum-rate maximization problem is formulated to optimize power and reflecting element allocation across UEs. Furthermore, we analyze and compare fully- and sub-connected active RIS architectures, revealing critical trade-offs between total throughput and energy consumption. Numerical results confirm that active RIS outperforms passive RIS systems in QoS aspects, and the choice of sub-connected configurations can provide a balanced compromise between performance and efficiency. These results position HAPS-active RIS as a promising enabler for scalable, energy-aware, and high-capacity NTN deployments in 6G networks. Future research can focus on practical phase shift design, energy efficiency optimization, and the integration of non-orthogonal multiple access~(NOMA) techniques to enhance user connectivity in HAPS-RIS-assisted communication systems.


\bibliographystyle{IEEEtran}
\bibliography{bibliography}

\begin{thebibliography}{10}
\providecommand{\url}[1]{#1}
\csname url@samestyle\endcsname
\providecommand{\newblock}{\relax}
\providecommand{\bibinfo}[2]{#2}
\providecommand{\BIBentrySTDinterwordspacing}{\spaceskip=0pt\relax}
\providecommand{\BIBentryALTinterwordstretchfactor}{4}
\providecommand{\BIBentryALTinterwordspacing}{\spaceskip=\fontdimen2\font plus
\BIBentryALTinterwordstretchfactor\fontdimen3\font minus \fontdimen4\font\relax}
\providecommand{\BIBforeignlanguage}[2]{{%
\expandafter\ifx\csname l@#1\endcsname\relax
\typeout{** WARNING: IEEEtran.bst: No hyphenation pattern has been}%
\typeout{** loaded for the language `#1'. Using the pattern for}%
\typeout{** the default language instead.}%
\else
\language=\csname l@#1\endcsname
\fi
#2}}
\providecommand{\BIBdecl}{\relax}
\BIBdecl

\bibitem{6Groad}
W.~Jiang, B.~Han, M.~A. Habibi, and H.~D. Schotten, ``The road towards 6\uppercase{G}: A comprehensive survey,'' \emph{IEEE Open J. Commun. Soc.}, vol.~2, pp. 334--366, 2021.

\bibitem{kement}
C.~E. Kement, F.~Kara, W.~Jaafar, H.~Yanikomeroglu, G.~Senarath, N.~D. Dào, and P.~Zhu, ``{Sustaining dynamic traffic in dense urban areas with high altitude platform stations (HAPS)},'' \emph{IEEE Commun. Mag.}, vol.~61, no.~7, pp. 150--156, Jul. 2023.

\bibitem{gunessurvey}
G.~K. Kurt, M.~G. Khoshkholgh, S.~Alfattani, A.~Ibrahim, T.~S.~J. Darwish, M.~S. Alam, H.~Yanikomeroglu, and A.~Yongacoglu, ``A vision and framework for the high altitude platform station (\uppercase{HAPS}) networks of the future,'' \emph{IEEE Commun. Surveys Tuts.}, vol.~23, no.~2, pp. 729--779, 2021.

\bibitem{safwanlinkbudget}
S.~Alfattani, W.~Jaafar, Y.~Hmamouche, H.~Yanikomeroglu, and A.~Yongaçoglu, ``Link budget analysis for reconfigurable smart surfaces in aerial platforms,'' \emph{IEEE Open J. Commun. Soc.}, vol.~2, pp. 1980--1995, 2021.

\bibitem{RIS1}
A.~L. Moustakas, G.~C. Alexandropoulos, and M.~Debbah, ``Reconfigurable intelligent surfaces and capacity optimization: A large system analysis,'' \emph{IEEE Trans. Wireless Commun.}, vol.~22, no.~12, pp. 8736--8750, 2023.

\bibitem{huang2019reconfigurable}
C.~Huang, A.~Zappone, G.~C. Alexandropoulos, M.~Debbah, and C.~Yuen, ``Reconfigurable intelligent surfaces for energy efficiency in wireless communication,'' \emph{IEEE Trans. Wireless Commun.}, vol.~18, no.~8, pp. 4157--4170, 2019.

\bibitem{safwanmagazine}
S.~Alfattani, W.~Jaafar, Y.~Hmamouche, H.~Yanikomeroglu, A.~Yongaçoglu, N.~D. Dào, and P.~Zhu, ``Aerial platforms with reconfigurable smart surfaces for \uppercase{5G} and beyond,'' \emph{IEEE Commun. Mag.}, vol.~59, no.~1, pp. 96--102, 2021.

\bibitem{safwanGlobecom}
S.~Alfattani, A.~Yadav, H.~Yanikomeroglu, and A.~Yongaçoglu, ``Beyond-cell communications via \uppercase{HAPS-RIS},'' in \emph{IEEE Globecom Workshops (GC Wkshps)}, 2022, pp. 1383--1388.

\bibitem{safwanLetter}
S.~\uppercase{A}lfattani, A.~Yadav, H.~Yanikomeroglu, and A.~Yongaçoglu, ``Resource-efficient \uppercase{HAPS-RIS} enabled beyond-cell communications,'' \emph{IEEE Wireless Commun. Lett.}, vol.~12, no.~4, pp. 679--683, 2023.

\bibitem{alfattani2022multi}
S.~Alfattani, W.~Jaafar, H.~Yanikomeroglu, and A.~Yonga{\c{c}}oglu, ``Multi-mode high altitude platform stations \uppercase{(HAPS)} for next generation wireless networks,'' \emph{arXiv preprint arXiv:2210.11423}, 2022.

\bibitem{azizi2024exploring}
A.~Azizi, M.~Kishk, and A.~Farhang, ``Exploring the impact of \uppercase{HAPS-RIS} on \uppercase{UAV}-based networks: A novel architectural approach,'' \emph{arXiv preprint arXiv:2409.17817}, 2024.

\bibitem{deka2024performance}
R.~Deka, M.~S. Alam, I.~Ahmed, and S.~Anees, ``Performance analysis of a \uppercase{RIS-HAPS} assisted \uppercase{FSO-UWOC} system for ground-air-underwater connectivity,'' in \emph{Proc. IEEE 100th Vehicular Technology Conference (VTC2024-Fall)}, 2024, pp. 1--5.

\bibitem{liu2021active}
K.~Liu, Z.~Zhang, L.~Dai, S.~Xu, and F.~Yang, ``Active reconfigurable intelligent surface: Fully-connected or sub-connected?'' \emph{IEEE Commun. Lett.}, vol.~26, no.~1, pp. 167--171, 2021.

\bibitem{yigit2022hybrid}
Z.~Yigit, E.~Basar, M.~Wen, and I.~Altunbas, ``Hybrid reflection modulation,'' \emph{IEEE Trans. Wireless Commun.}, vol.~22, no.~6, pp. 4106--4116, 2022.

\bibitem{3GPP_UE}
{3GPP}, ``{Study on new radio (NR) to support non-terrestrial networks},'' Tech. Rep. 38.863 38.811 Release 15, v15.4.0, Sept. 2020.

\bibitem{discreteLoss}
M.~Rivera, M.~Chegini, W.~Jaafar, S.~Alfattani, and H.~Yanikomeroglu, ``Optimization of quantized phase shifts for reconfigurable smart surfaces assisted communications,'' in \emph{Proc. IEEE 19th Annu. Consum. Commun. Netw. Conf. (CCNC)}, 2022, pp. 901--904.

\bibitem{boyd2007tutorial}
S.~Boyd, S.-J. Kim, L.~Vandenberghe, and A.~Hassibi, ``A tutorial on geometric programming,'' \emph{Optim. Eng.}, vol.~8, pp. 67--127, 2007.

\bibitem{khennoufa2024multi}
F.~Khennoufa, K.~Abdellatif, H.~Yanikomeroglu, M.~Ozturk, T.~Elganimi, F.~Kara, and K.~Rabie, ``Multi-layer network formation through \uppercase{HAPS} base station and transmissive \uppercase{RIS}-equipped \uppercase{UAV},'' \emph{arXiv preprint arXiv:2405.01692}, 2024.

\end{thebibliography}

\vfill

\end{document}